\documentclass[12pt]{article}
\usepackage{amssymb}\usepackage{bm}\usepackage{amsfonts}\usepackage{amsmath}
\usepackage{undertilde}\usepackage{hyperref}
\newcommand{\beq}{\begin{equation}}\newcommand{\eeq}{\end{equation}}\newcommand{\beqa}{\begin{eqnarray}}
\newcommand{\eeqa}{\end{eqnarray}}\newcommand{\w}{\wedge}\newcommand{\ts}{\textstyle}
\newcommand{\nn}{\nonumber}\newcommand{\vep}{\varepsilon}
\newcommand{\bep}{\bar{\varepsilon}}

\newcommand{\Lie}[1]{\mathcal{L}_{\bar{#1}}}

\def\p{\partial}

\begin{document}
{\renewcommand{\thefootnote}{\fnsymbol{footnote}}
\hfill  IGC--09/6--1\\
\medskip
\begin{center}
{\LARGE  Do Spinors Frame-Drag?}\\
\vspace{1.5em}
Andrew Randono\footnote{e-mail address: {\tt arandono@perimeterinstitute.ca}}
\\
\vspace{0.5em}
Institute for Gravitation and the Cosmos,\\
The Pennsylvania State
University,\\
104 Davey Lab, University Park, PA 16802, USA\\
\   \\
and \\
\   \\
The Perimeter Institute for Theoretical Physics \\
31 Caroline Street North\\
Waterloo, ON N2L 2Y5, Canada
\vspace{1.5em}
\end{center}
}

\setcounter{footnote}{0}

\begin{abstract}
We investigate the effect of the intrinsic spin of a fundamental spinor field on the surrounding spacetime geometry. We show that despite the lack of a rotating stress-energy source (and despite claims to the contrary)  the intrinsic spin of a spin-half fermion gives rise to a frame-dragging effect analogous to that of orbital angular momentum, even in Einstein-Hilbert gravity where torsion is constrained to be zero. This resolves a paradox regarding the counter-force needed to restore Newton's third law in the well-known spin-orbit interaction. In addition, the frame-dragging effect gives rise to a {\it long-range} gravitationally mediated spin-spin dipole interaction coupling the {\it internal} spins of two sources. We argue that despite the weakness of the interaction, the spin-spin interaction will dominate over the ordinary inverse square Newtonian interaction in any process of sufficiently high energy for quantum field theoretical effects to be non-negligible. 
\end{abstract}


\section{Introduction}
The Einstein field equations bear the surprising consequence that rotating mass distributions tend to drag their local spacetime frame with them as they rotate. This well-known effect gives rise to the precession of spinning test masses as well as a gravitationally mediated dipole-dipole force (see e.g. \cite{WheelerCiufolini}). It has been known for almost a century now that the total angular momentum of a fundamental field, such as a Dirac field, is not determined by its orbital angular momentum alone, but only the combination of orbital and spin angular momentum. In this paper, we address a simple question: {\it Do spinors frame-drag?} Or, more specifically, {\it does the internal spin of a spinor give rise to frame-dragging effects analogous to that of orbital angular momentum?} The caveat in the case of internal spin is of course that nothing is actually rotating in spacetime, since spin is an abstract, internal property of fundamental particles.

Let us first draw attention to a hole in the current understanding of the coupling of intrinsic spin to the gravitational field. For some time it has been known that in addition to the standard gravitationally mediated orbit-orbit interaction, there is a gravitationally mediated spin-orbit dipole interaction as pointed out by Mashhoon \cite{Mashhoon--2008,Mashhoon--Gravitomagnetism,MashhoonKaiser}, Hehl and Ni \cite{HehlNi}, Obukhov \cite{Obukhov}, and others. The existence of such an interaction follows from a straight forward application of the equivalence principle to rotating frames of reference. Consider then the dipole-dipole force between a rotating massive object and the intrinsic spin of a Dirac field. As argued from the equivalence principle, the force on the spinor will be opposite to the electromagnetic dipole interaction in that north-pole attracts north-pole and south-pole attracts south-pole, a characteristic first noticed by Wald in the context of a spinning body rotating around a spinning blackhole \cite{Wald:Spin}. The details of the calculation reveal that the force is mediated by the gravimagnetic field produced by the rotating mass distribution. On the other hand, for Newton's third law to hold in the static case (static so that momentum is not carried by the field itself), there must be an equal and opposite force exerted by the spinor on the rotating mass distribution. But, what field mediates this interaction? Is it mediated by a gravimagnetic field sourced by the intrinsic spin itself? Does intrinsic spin frame-drag? Such a frame-dragging effect has been demonstrated for the specific case of a rigid spherical and cylindrical spin-polarized source \cite{Arkuszewski, Tsoubelis--framedrag1,Tsoubelis--framedrag2}, however to our knowledge it has not been properly derived from an arbitrary source consisting of a Dirac spinor field. If such a source does produce a frame-dragging effect, then it leaves open the possibility of a {\it spin-spin} dipole interaction. This interaction is distinct from the contact interaction mediated by nonpropagating torsion in Einstein-Cartan theory, since it is expected to be a {\it long-distance} interaction\footnote{By long distance here, we simply mean that it is not a contact interaction. General principles suggest it should be a dipole-dipole interaction so the force should fall off like $r^{-4}$. This is considered short distance by many comparisons, but it is {\it not} a contact interaction.} coupling the intrinsic spins of two spatially isolated spinor fields. 

In this paper we address these questions in the linearized theory of tetrad gravity coupled to a spinor field. In particular, in the context of linearized gravity, we show that the intrinsic spin angular momentum, or more specifically, the expectation value of intrinsic spin, gives rise to a frame-dragging effect exactly analogous to its orbital angular momentum. Thus, the total source for the gravimagnetic field is the net angular momentum consisting of intrinsic spin plus orbital angular momentum. The gravitational field at asymptotic infinity must then contain information about both the spin and orbital angular momentum, and we show that our recently proposed generalization of the the total angular momentum to include intrinsic spin as a surface integral at asymptotic infinity \cite{RandonoSloan} does in fact give the right answer in this context. Furthermore, under the magneto-static approximation it has the familiar functional form as the analogous expression for the ordinary magnetic moment as a surface integral following from a straightforward application of Stoke's theorem.

We then show directly from the gravitational Hamiltonian of an asymptotically flat, static  spacetime that, in addition to the orbit-orbit and spin-orbit interaction terms of two isolated spinor sources, there is, indeed, a gravitationally mediated spin-spin interaction. This is in contrast to what appears to have been the prevailing belief that such an interaction could not occur without propagating torsion. In this case, the interaction is mediated by the ordinary gravitational tetrad even when the torsion is constrained to be identically zero. 

The spin-spin interaction has the same functional form as the analogous electromagnetic dipole-dipole interaction, but with Newton's constant playing the role of $1/4\pi \epsilon_0$, and with an opposite sign. Since the gravitational interaction is extremely weak already, and the dipole-dipole force falls off like $r^{-4}$ as opposed to the $r^{-2}$ fall-off of the ordinary Newtonian interaction, it may be argued that in any experimentally realizable scenario the spin-spin interaction will be negligibly small as compared to the ordinary Newtonian interaction. However, since the spin to mass ratio of fundamental particles is rather large (as compared to, for example, the extremal limit of a Kerr black hole), from generic arguments we will show that the spin-spin interaction is expected to dominate at length scales smaller than the geometric mean of the Compton wavelengths of the two interacting particles. Thus, as a general rule of thumb, we claim that the $r^{-4}$ spin-spin interaction will dominate over the $r^{-2}$ Newtonian interaction in any process of sufficiently high energy for quantum field theoretic effects to play a significant role. 

Our conventions are as follows. The Lorentzian metric signature is $\{-,+,+,+\}$. Lower case Greek indices $\{\alpha, \beta,\mu,\nu,...\}$ represent spacetime indices in the tangent space of the four-- dimensional spacetime manifold. Lower-case Roman indices near the beginning of the alphabet, $\{a,b,c,...\}$, are spatial indices in the tangent space of a three--dimensional spatial hypersurface. Upper-case Roman indices $\{I,J,K,...\}$ are indices in a vector representation of the internal $Spin(3,1)$ gauge group, and lower case Roman indices near the middle of the alphabet, $\{i,j,k,...\}$, are spatial indices in the fiber. Since we are working in the opposite signature from most quantum field theory texts, the Dirac inner product must be defined with an extra $i$ factor so that $\bar{\psi}=i\,\psi^{\dagger} \gamma^0$. There are different conventions in the literature for what the words ``orbit" and ``spin" mean. In this paper, the word {\it spin} will always refer to the internal spin angular momentum of a spin-$\frac{1}{2}$ field, and the term {\it orbit} will always refer to the rotational motion of, say, a rigid rotating object.
 
\section{Linearized Gravity in Tetrad Variables}
In order to analyze the geometric consequences of spin in general relativity, it is most convenient to consider the linearized theory where many basic concepts such as angular momentum and dipole moment that are familiar from flat space theories carry over to curved spacetime. In order to couple spinors to the gravitational field, it is necessary to resort to the tetrad formulation of gravity where the dynamical metric, $g_{\mu\nu}$ is replaced by the dynamical tetrad ${e^I}_\mu $ related to the metric by $g_{\mu\nu}=\eta_{IJ}{e^I}_\mu\,{e^J}_\nu$, where $\eta_{IJ}=diag(-1,1,1,1)$ is the canonical bilinear form. In the linearized theory, the tetrad is expanded about a fiducial flat metric $^0{e^I}_\mu$, which for convenience we will set to be the canonical tetrad $^0{e^I}_\mu={\delta^I}_\mu$ so that the tetrad and its inverse become
\beq
{e^I}_\mu={\delta^I}_\mu+{\vep^I}_\mu \quad \quad {e^\nu}_J={\delta^\nu}_J-{\vep^\nu}_J\,.
\eeq
The perturbations $\vep$ are assumed to be small enough that we consider only expressions that are first order in $\vep$. By convention, in the linearized theory, we will raise (lower) indices using the flat metrics $\eta^{IJ}(\eta_{KL})$ and $\eta^{\alpha\beta}(\eta_{\mu\nu})$ and convert between internal and spacetime indices with ${\delta^I}_\mu({\delta^\nu}_J)$. Ordinarily the index placement of $e^I_\mu$ is immaterial since the object is itself the map between the tangent space and the internal $SO(3,1)$ vector space -- however, since the fiducial flat tetrad now plays that role, the index placement of ${\vep^I}_\mu$ does matter. Specifically, the relation between $\vep$ and the ordinary metric perturbations is $h_{\mu\nu}=2\,\vep_{(\mu\nu)}$ where $g_{\mu\nu}=\eta_{\mu\nu}+h_{\mu\nu}$ and following our conventions, $\vep_{\mu\nu}=\eta_{IJ}{\delta^I}_\mu {\vep^J}_\nu$. We will adopt the convention that antisymmetric component fields will be denoted with an underbar, and symmetric components will be denoted with an undertilde so that ${\vep^I}_\mu={\underline{\vep}^I}_\mu+{\utilde{\vep}^I}_\mu$ with ${\underline{\vep}^I}_\mu=\eta^{I\alpha}\vep_{[\alpha\mu]}$ and ${\utilde{\vep}^I}_\mu=\eta^{I\alpha}\vep_{(\alpha\mu)}$.

Since we wish to emphasize that a spin-spin interaction arises even without a propagating torsional field, we will work in the second order formulation of Einstein-Hilbert gravity formulated in terms of tetrads. For more on the low-energy effects of torsional theories, see for example \cite{SinghRyder, Hammond--Torsion}. Given a basis of coframes, the torsion-free spin-connection coefficients can be written
\beq
{\Gamma^{IJ}}_\mu[e]=2{e^{[I|\rho|}}{e^{J]}}_{[\rho,\mu]}+e^{I\alpha}e^{J\beta}e_{M\mu}{e^M}_{[\alpha,\beta]}
\eeq
which in the linearized theory reduces to
\beq
{\Gamma^{IJ}}_\mu=2\eta^{\nu [I}{\vep^{J]}}_{[\nu,\mu]}+\eta_{\mu K}\eta^{\nu I} \eta^{\sigma J} {\vep^K}_{[\nu,\sigma]}\,.
\eeq
Writing the curvature ${R^{IJ}}_{\mu\nu}=2 (\p_{[\mu}{\Gamma^{IJ}}_{\nu]} +{\Gamma^I}_{K[\mu}{\Gamma^{KJ}}_{\nu]})$ and the Ricci tensor ${R^I}_\mu={e^\nu}_J\,{R^{IJ}}_{\mu\nu}$, the Einstein equations are 
\beq
{G^I}_\mu={R^I}_\mu-\frac{1}{2}{e^I}_\mu R=8\pi G \ {\tau^I}_\mu\,.
\eeq

As usual, by choosing a gauge in which the divergence of the perturbation of the tetrad vanishes, ${\vep^I}{}_\sigma{}^{,\sigma}=0$, the linearized Einstein equations take the form
\beq
-\Box \,{\utilde{\bep}^I}{}_\mu =16\pi G \,{\tau^I}_\mu\,.
\eeq
where  ${\utilde{\bep}^I}{}_\mu = {\utilde{\vep}}^I {}_\mu-{\ts \frac{1}{2}}\,\delta^I_{\mu}\, \vep $ and $\vep = \delta_I^\mu \,{\vep^I}_\mu$. 

\section{Fermionic Matter Sources}
We will now take the source to be a Dirac field. The total action for the system is 
\beqa
S&=& S_g +S_\psi \\
&=& \frac{1}{16\pi G}\int d^4x\,e\,\left(\, {e^\mu}_I {e^\nu}_J \, {R[\Gamma]^{IJ}}_{\mu\nu} \right)\\
& & +\alpha\int d^4 x\, e \,\left(\frac{1}{2}{e^\mu}_I \left(\bar{\psi}\gamma^I (D_{\mu}\psi)-(D_\mu \bar{\psi}) \gamma^I \psi \right)-m\bar{\psi}\psi \right)\,.
\eeqa
Here we have included a coupling constant $\alpha$ to account for different conventions. Normally, the constant can be taken to be $\alpha=-1$.

Variation with respect to the spinor field gives the Dirac equation
\beq
{e^\mu}_I\gamma^I D_\mu \psi=m\psi 
\eeq
where $D_\mu\psi = \p_\mu\psi+\frac{1}{4}\gamma_{[I}\gamma_{J]}\,{\Gamma^{IJ}}_\mu\,\psi$.
Variation with respect to the tetrad yields the Einstein-Cartan equations:
\beq
{R^I}_\mu-\frac{1}{2}{e^I}_{\mu} R=8\pi G {\tau^I}_\mu
\eeq
where the stress-energy is given by
\beq
{\tau^I}_\mu=-\alpha\,e^{I\alpha}\,\frac{1}{2}\left(\bar{\psi}\gamma_{(\alpha} D_{\mu)}\psi -D_{(\mu}\bar{\psi}\gamma_{\alpha)}\psi \right)\,.
\eeq
For future reference, it will be useful to mention the following identity which follows from the extra six components of the fixed points of the variation of the action with respect to ${e^I}_\mu$, but also follows from the Dirac equation:
\beq
\frac{1}{2}\left(\bar{\psi}\gamma_{[\alpha} D_{\mu]}\psi -D_{[\mu}\bar{\psi}\gamma_{\alpha]}\psi \right)=-\frac{i}{4}{\epsilon_{\alpha\mu}}^{\sigma\rho}D_\sigma J^{(5)}_{\rho} \label{Anti-Identity}
\eeq
where $J^{(5)}_{\rho}=\bar{\psi}\gamma_5 \gamma_\rho \psi$ is the axial current.

\section{Multipole Expansion of the vector potential}
In this section we review the generic procedure for calculating the vector potential from an current distribution in the monopole-dipole approximation (for the electromagnetic case, see e.g. \cite{Jackson:EM}). The generic wave equation 
\beq
-\Box A^{\mu}=\kappa J^\mu
\eeq
can be inverted using retarded Green's functions to yield
\beq
A^\mu(x)=\frac{\kappa}{4\pi}\int_{\Lambda_{-}}d^3x'\frac{J^\mu(x')}{\mid\bm{r}-\bm{r'} \mid}
\eeq
where the integral is over the past light cone. Under the assumption that the current distribution is localized near the origin, one can expand the integrand in powers of $r$ so that $|\bm{r}-\bm{r'}|^{-1}=1/r+\bm{\hat{r}} \cdot \bm{r'} /r^2$, yielding
\beq
A^\mu(x)=\frac{\kappa}{4\pi}\left(\frac{1}{r}\int_{\Lambda_-}d^3 x'\,J^\mu(x')+\frac{\hat{r}_\nu}{r^2}\int_{\Lambda_-}d^3 x'\, x'^\nu J^\mu(x')\,d^3 x'\right)\,, \label{Dipole-Monopole}
\eeq
where $r^\mu=x^\mu-\delta^\mu_0\,x^0$ is the spatial radial vector and $\hat{r}^\mu=r^\mu/r$. Under the assumption that the current is localized and conserved, $\p_\mu J^\mu=0$, one can derive the following identities:
\beqa
\int_{\Lambda_-}d^3 x'\,J^\mu &=&\delta^\mu_0\int_{\Lambda_-}d^3 x'\,J^0+\int_{\Lambda_-}d^3 x'\,x'^\mu\p_t J^0 \\
\int_{\Lambda_-}d^3 x'\,\left(x'^{(\alpha} J^{\beta)}\right)&=& \int_{\Lambda_-}d^3 x'\, \delta^{(\alpha}_0 x'^{\beta)}\,J^0 +\int_{\Lambda_-}d^3 x'\,x'^\alpha x'^\beta\,\p_t J^0\,.
\eeqa
To simplify the expressions, we make the standard ``magnetostatic" assumption that the charge distribution is conserved in time so that we have the additional constraint
\beq
\p_t J^0=0\quad\quad \p_a J^a=0 \,.
\eeq
Under this assumption, the field distribution is static so the integral over the past light cone can be replaced by an integral over a constant time slice. Then, by splitting the second integral on the right--hand side of (\ref{Dipole-Monopole}) into its symmetric and antisymmetric components, the expression reduces to
\beq
A^\mu=\frac{\kappa}{4\pi}\left( \delta^\mu_0 \left(\frac{Q}{r}+\frac{\bm{\hat{r}}\cdot {\ts\frac{1}{2}}\bm{p}}{r^2}\right)-\frac{\hat{r}_\nu \,M^{\mu\nu}}{r^2}\right) \label{VectorExpansion}
\eeq
where
\beq
Q\equiv \int_{\Sigma}d^3 x'\,J^0
\eeq
is the total charge, 
\beq
{\bm{p}} \equiv \int_{\Sigma} d^3 x' \, \, {\bm{r'}} J^0
\eeq
is the electric dipole moment\footnote{As we have written expression (\ref{VectorExpansion}), half of the electric dipole momentum appears in the relativistic angular momentum, explaining the explicit $\frac{1}{2}$ in the expression.}, and 
\beq
{M}^{\mu\nu}\equiv \int_{\Sigma}d^3 x'\,x'^{[\mu} J^{\nu]}
\eeq
is the relativistic generalization of the magnetic moment whose spatial components are related to the ordinary magnetic moment by $({\bm{m}})^a=\frac{1}{2}{\epsilon^a}_{bc}M^{bc}$. Splitting this into spatial and time components, we have
\beqa
A^0 &=& \frac{\kappa}{4\pi}\left(\frac{Q}{r}+\frac{\bm{\hat{r}\cdot p}}{r^2}\right) \\
A^a &=& \frac{\kappa}{4\pi}\frac{\hat{r}_b M^{ba}}{r^2} 
\eeqa
Focusing on the spatial components of the vector potential alone, we rewrite the above in vector notation as
\beq
\bm{A}=\frac{\kappa}{4\pi}\frac{\bm{m}\times \bm{\hat{r}}}{r^2}\,,
\eeq
which gives rise to the dipole magnetic field
\beq
\bm{B}=\bm{\nabla}\times \bm{A}=\frac{\kappa}{4\pi}\left[\frac{3\bm{\hat{r}}(\bm{\hat{r}}\cdot \bm{m})-\bm{m}}{|\bm{x}|^3}+\frac{8\pi}{3} \bm{m}\,\delta^3(\bm{r})\right] \,.\label{DipoleField}
\eeq
The delta-distribution term in the magnetic field is the well-known Fermi contact term that gives rise to the hyperfine interaction in the case of ordinary electromagnetism. For future reference, it is required so that a surface integral expression for the dipole moment agrees with a bulk integral expression. Specifically, given the form of the vector potential $\bm{A}$, it is easy to see that
\beq
\frac{3}{2\mu_0}\int_{S^2}\bm{\hat{r}}\times\bm{A} \,\,R^2 d\Omega =\bm{m}\,.
\eeq
Using Stoke's theorem to yield $\int_{r<R} \bm{\nabla \times A} \,\,d^3 x= R^2 \int \bm{\hat{r}\times A} \,\,d\Omega $, we see that the bulk and the surface integrals only agree if the delta-distribution is included in the magnetic field.

\subsection{The magnetic dipole field of a spinor}
To derive the magnetic dipole field of a spinor, we take the vector potential $A^\mu$ to be the ordinary electromagnetic vector potential so that $\kappa=\mu_0$, and we take the current to be the electromagnetic current of a massive Dirac field with charge $q$:
\beq
J^\mu=i q\, j^\mu_{Dirac}=iq\,\bar{\psi}\gamma^\mu\psi \,.
\eeq
We first notice that subject to the Dirac equation, $\gamma^\mu D_\mu \psi =m\psi$ where $D_\mu=\p_\mu-i q A_\mu$, the Dirac current splits into two, separately conserved currents:
\beqa
j^\mu_{Dirac}&=&j^\mu_{KG}+j^\mu_{spin}\\
&=& \frac{i}{2m}\left(\bar{\psi}D^\mu\psi -(D^\mu\bar{\psi})\psi \right)+\frac{i}{4m}\p_\alpha \left(\bar{\psi}[\gamma^\mu,\gamma^\alpha]\psi\right)\,.
\eeqa
This splitting is often referred to as the Gordon decomposition of the current. We emphasize this step because a similar decomposition will allow us to separate the orbital and spin components of the corresponding gravitational current, the stress-energy. Using this identity, upon integration by parts the relativistic magnetic moment becomes
\beq
M^{\mu\nu}=q\int_{\Sigma}d^3 x'\,x'^{[\mu} j^{\nu]}_{KG}+\frac{iq}{4m}\int_{\Sigma}d^3 x'\,\bar{\psi}[\gamma^\mu,\gamma^\nu]\psi\,.
\eeq
Note that the term coming from the spin current contains no derivatives, so it depends only on intrinsic spin properties of the Dirac field. As usual, to examine the nonrelativistic limit of the Dirac equation, we work in a representation where the Dirac matrices take the form:
\beq
\gamma^0=-i\left[\begin{array}{cc}1& 0 \\ 0&-1 \end{array} \right] \quad \quad 
\gamma^i=-i\left[\begin{array}{cc}0& \sigma^i \\ -\sigma^i &0 \end{array} \right] \quad \quad 
\gamma_5=\left[\begin{array}{cc}0& 1 \\ 1&0 \end{array} \right]\,.
\eeq
In the nonrelativistic limit, the Dirac field splits into large and small components:
\beq
\psi=\left[\begin{array}{c} \Phi \\ \phi \end{array}\right] 
\eeq
and the large components dominate, so that, for example, the relevant currents can be approximated by 
\beq
i\bar{\psi}\gamma^0 \psi \approx \Phi^{\dagger} \Phi \quad i\bar{\psi}[\gamma^{i},\gamma^j]\psi \approx -2{\epsilon^{ij}}_k \Phi^{\dagger}\sigma^k \Phi  \quad i\bar{\psi}\gamma_5 \gamma^i \psi \approx -\Phi^\dagger \sigma^i\Phi \,.
\eeq
In this limit, the magnetic moment becomes
\beq
\frac{1}{2}\epsilon^{abc}M_{bc}=\frac{-q}{2m}\left[\int_\Sigma d^3 x'\,\frac{-i}{2}\epsilon^{abc}\left(\Phi^{\dagger}x_b D_c \Phi -(x_b D_c \Phi^{\dagger})\Phi \right) +2\times \int_\Sigma d^3 x'\,\frac{1}{2}\Phi^\dagger \sigma^a \Phi \right]\,.
\eeq
To understand this better, we compare this result to the Noether charge associated with angular momentum given by (see e.g. \cite{RandonoSloan})
\beq
Q^{\{\hat{I}\hat{J}\}}=\int_\Sigma d^3 x\, \delta^{\hat{I}\hat{J}}_{\mu\nu}\left(\frac{1}{2}\left(\bar{\psi}\gamma^0 x^\mu D^\nu \psi - (x^\mu D^\nu \bar{\psi})\gamma^0 \psi\right) +\frac{i}{4}\epsilon^{0\mu\nu\alpha}\bar{\psi}\gamma_5\gamma_\alpha\psi \right)\,.
\eeq
The nonrelativitic limit of this expression yields the angular momentum expression:
\beqa
(\bm{J_{tot}})^a &=& (\bm{L+S})^a = \frac{1}{2}{\epsilon^a}_{bc}Q^{\{bc\}}\nn\\
&\approx &\int d^3 x\, \frac{-i}{2}{\epsilon^a}_{bc}\left(\Phi^{\dagger} x^b D^c \Phi -(x^b D^c \Phi^\dagger ) \Phi \right) +\int d^3 x\,\frac{1}{2}\Phi^\dagger \sigma^a \Phi \,.
\eeqa
The first integral is clearly the expectation value of the orbital angular momentum, and the second is the expectation value of the spin. Comparing these expressions, we obtain the expected form for the magnetic moment:
\beq
\bm{m}_{EM}=\frac{-q}{2m}\left(\bm{L}+2\bm{S}\right)\,.
\eeq
It is therefore clear from the discussion leading up to (\ref{DipoleField}) that the internal spin of a localized Dirac field produces its own dipole magnetic field characterized by a dipole moment of twice the spin angular momentum. 

For the gravitational interaction, the procedure is very much analogous to the above derivation. In this case, the linearized perturbations to the tetrad will play the role of the gauge field, which is sourced by the stress-energy as the conserved current. A corresponding ``gravitational Gordon decomposition" of the stress-energy will allow a separation of orbital angular momentum and spin. In place of the Dirac bilinear $\bar{\psi}\gamma^{[\mu}\gamma^{\nu]}\psi$ as the intrinsic spin source for the magnetic field, the axial current, $\bar{\psi}\gamma_5\gamma^\mu \psi$, will play the role of the intrinsic spin source for the gravimagnetic field.

\section{The dipole expansion of the gravimagnetic field}
The perturbation to the tetrad in the linearized theory satisfies the same set of equations as the vector potential given above. That is, we can think of ${\utilde{\bep}^I}{}_\mu$ as a set of four vector potentials labeled by the internal index $I$. The coupling constant is $\kappa=8\pi G$. Thus we have the immediate solution:
\beq
{\utilde{\bep}^I}{}_\mu=2G \left( \eta_{\mu 0} \left(\frac{Q^I}{r}+\frac{\bm{\hat{r}}\cdot {\ts\frac{1}{2}}\bm{p}^I}{r^2}\right)-\frac{\hat{r}^\nu \,M^I{}_{\mu\nu}}{r^2}\right)
\eeq
or separating space and time components,
\beqa
{\utilde{\bep}^I}{}_0 &=& -2G\, \left( \frac{Q^I}{r} + \frac{\bm{\hat{r}}\cdot \bm{p}^I}{r^2}\right) \\
{\utilde{\bep}^I}{}_a &=& 2G\, \frac{\hat{r}^b M^I{}_{ba}}{r^2} 
\eeqa
where by analogy we have defined
\beqa
Q^I &=& \int_\Sigma d^3 x'\, {\tau^{I}}_\alpha \,\eta^{\alpha 0} \nn\\
\bm{p}^I &=&  \int_\Sigma d^3 x'\, \bm{x'} {\tau^{I}}_\alpha \,\eta^{\alpha 0} \nn\\
M^I_{\mu\nu} &=& \int_\Sigma d^3 x'\, x'_{[\mu}{\tau^I}_{\nu]}\,.
\eeqa

Let us now focus on the time components ${\utilde{\bep}^{\hat{0}}}_\mu$. Analogous to ordinary gravitomagnetism, in a low energy, low velocity limit we expect these terms to dominate. Since the resulting field equations in this gauge are very similar to the electromagnetic field equations in the Lorentz gauge, as in ordinary gravitomagnetism we define the gravito-electromagnetic vector potential $\mathcal{A}_\mu\equiv {\utilde{\bep}^{\hat{0}}}_\mu$. In the nonrelativistic limit, the charge $Q^{\hat{0}}$ reduces to the expectation value of the energy:
\beq
Q^{\hat{0}}\approx -\alpha \langle \hat{E} \rangle= -\alpha \int_{\Sigma}d^3 x' \frac{1}{2}\left(\Phi^{\dagger} i\partial_t \Phi -(i\partial_t \Phi^{\dagger})\Phi \right)
\eeq
and the dipole $\bm{p}^{\hat{0}}$ is the dipole associated with the energy density.

We now focus on the magnetic moment term $M^{\hat{0}}_{\mu\nu}$. As in the case of electromagnetism, the first step is to rewrite the matter current so that the internal spin of the particle is more transparent in the current. To do this, we employ the identity (\ref{Anti-Identity}), which allows us to write the stress tensor in a more convenient form analogous to the Gordon decomposition of the Dirac current (for a more in-depth review of the Gordon decomposition of the gravitational stress-energy and spin-currents, see \cite{HehlRyder:Gordon}):
\beq
{\tau^I}_\mu=-\alpha\left(\frac{1}{2}\left(\bar{\psi}\gamma^I D_\mu \psi -D_\mu \bar{\psi}\gamma^I \psi\right)+\frac{i}{4}{\epsilon^I{}_\mu}^{\alpha\beta} D_\alpha J^{(5)}_\beta \right)
\eeq
In the linearized limit, the components $M^{\hat{0}}_{\mu\nu}$ in the magnetic dipole contribution of the multipole expansion reduce to
\beq
M^{\hat{0}}_{\mu\nu}=-\alpha \int_\Sigma d^3 x' \frac{1}{2} \left( \bar{\psi}\gamma^0 x_{[\mu}\p_{\nu]}\psi -(x_{[\mu}\p_{\nu]}\bar{\psi})\gamma^0 \psi \right)-\frac{i}{4} {\epsilon_{0\mu\nu}}^\sigma \bar{\psi}\gamma_5\gamma_\sigma\psi \,.
\eeq
We recognize this as proportional to the conserved Noether charge associated with angular momentum: 
\beq
M^{\hat{0}}_{\mu\nu}=-\frac{\alpha}{2} Q_{\{\mu\nu\}}\,.
\eeq
Thus, from our previous calculation, we see that the spatial components of the gravimagnetic moment, $(\bm{m_g})^a=\frac{1}{2}{\epsilon^a}_{bc}M^{bc}$, are precisely one--half of the total  angular momentum:
\beq
\bm{m_g}=-\frac{\alpha}{2}(\bm{L}+\bm{S})\,.
\eeq
The resulting gravimagnetic vector potential has the form of a dipole potential,
\beq
\bm{\mathcal{A}}=2G\frac{\bm{m_g}\times \bm{\hat{r}}}{r^2}\,,\label{DipoleGPotential}
\eeq
which gives rise to the dipole gravimagnetic field
\beq
\bm{\mathcal{B}}=\bm{\nabla}\times \bm{\mathcal{A}}=2G\left[\frac{3\bm{\hat{r}}(\bm{\hat{r}}\cdot \bm{m_g})-\bm{m_g}}{r^3}+\frac{8\pi}{3} \bm{m_g}\,\delta^3(\bm{x})\right] \,.\label{DipoleGField}
\eeq
In the special case where the particle is localized and at rest, so that the angular momentum vanishes, the spin of the particle itself produces the gravimagnetic dipole field:
\beq
\bm{\mathcal{B}}^{spin}=-G\alpha \left[\frac{3\bm{\hat{r}}(\bm{\hat{r}}\cdot \bm{S})-\bm{S}}{r^3}+\frac{8\pi}{3} \bm{S}\,\delta^3(\bm{x})\right] \,.\label{GSpinDipole}
\eeq

\section{Computing the internal spin angular momentum in the linearized limit}
We recall that in magnetostatics, the magnetic moment of a system can be computed by a surface integral at infinity as follows:
\beq
\bm{m}=\lim_{r \to \infty} \frac{3}{2\kappa} \int_{S^2}r^2 d\Omega \,\,\bm{\hat{r}\times A}\,. \label{SurfaceIntegral}
\eeq
Since the monopole contribution is zero, and the higher order terms in the multipole expansion fall off faster than $r^{-2}$, the integral picks up precisely the dipole contribution of the magnetic field. In a recent paper, we have constructed a similar expression for the total (spin + orbital) angular momentum of an asymptotically flat spacetime in the tetrad framework, and related it to the Noether charge corresponding to angular momentum in the axisymmetric case. We will now show in the linearized limit, that this integral gives the analogue of (\ref{SurfaceIntegral}) for the gravimagnetic field.

The expression for the total angular momentum derived in \cite{RandonoSloan} is the following (here we have normalized the expression in \cite{RandonoSloan} by an overall minus sign):
\beqa
J^{\hat{I}\hat{J}}_{tot}& =& L^{\hat{I}\hat{J}}+S^{\hat{I}\hat{J}}   \\
L^{\hat{I}\hat{J}}&= &\frac{1}{4k}\int_\Sigma \epsilon_{IJKL}\,\iota_{\bar{K}^{\{\hat{I}\hat{J}\}}}(e^I \w e^J)\w \omega^{KL} \\
S^{\hat{I}\hat{J}}&=&\frac{1}{4k}\int_\Sigma{\epsilon^{IJ}}_{KL}\lambda^{\{\hat{I}\hat{J}\}}_{IJ} e^K\w e^L\,.
\eeqa
In the above expression, $k=8\pi G$, and the vectors $\bar{K}^{\{\hat{I}\hat{J}\}}$ form a six--dimensional basis for the boost and rotation Killing vectors of the fiducial flat metric ${}^0 g_{\mu\nu}$; the internal boost and rotation parameters $\lambda^{\{\hat{I}\hat{J}\}}_{IJ}$ are related to Killing vectors by
\beq
\mathcal{L}_{\bar{K}^{\{\hat{I}\hat{J}\}}}{}^0e^I = - (\lambda^{\{\hat{I}\hat{J}\}})^I{}_K\,{}^0 e^K\,.
\eeq
In the Cartesian basis where ${}^0 e^I{}_{\mu}=\delta^I{}_\mu$, the Killing basis vectors can be written $K^\mu_{\{\hat{I}\hat{J}\}}=-\lambda^{\mu\nu}_{\{\hat{I}\hat{J}\}}\,x_\nu$ and the internal boost and rotation parameters are given by $\lambda^{IJ}_{\{\hat{I}\hat{J}\}}= \delta^I_\mu \delta^J_\nu \,\lambda^{\mu\nu}_{\{\hat{I}\hat{J}\}} =\delta^I_{\hat{I}} \delta^J_{\hat{J}} -\delta^I_{\hat{J}} \delta^J_{\hat{I}}$. In addition to this, one also has the linear momentum given by (again correcting by an overall minus sign)
\beq
P^{\hat{I}}=\frac{1}{4k}\int_\Sigma \epsilon_{IJKL}\,\iota_{\bar{K}^{\{\hat{I}\}}} (e^I \w e^J)\w \omega^{KL} 
\eeq
where $\bar{K}^{\{\hat{I}\}}$ is a translational Killing vector for the fiducial flat metric given in the Cartesian basis by $\bar{K}^{\{\hat{I}\}}=\delta^{\hat{I}}_\mu \frac{\p}{\p x_\mu}$.

The key property of the spin and orbital angular momentum generators is that they satisfy the algebra of the spin enlarged Poincar\'{e} group under the Poisson bracket. The spin enlarged Poincar\'{e} algebra is the subgroup, $\mathfrak{G}({}^0 e)\subset Spin(3,1)\rtimes Diff_4$, of the local gauge group of Einstein-Cartan gravity that preserves the fiducial flat tetrad. The group has the structure of $Spin(3,1) \otimes (SO(3,1)\ltimes \mathbb{R}^{3,1})$ together with a two-to-one mapping between the $Spin(3,1)$ and $SO(3,1)$ subgroups that ensures that the group preserves ${}^0 e$.

We will now evaluate this expression for the spatial angular momentum components $J^{\hat{i}}_{tot}=\frac{1}{2}{\epsilon^{\hat{i}}}_{\hat{j}\hat{k}}\,J^{\hat{j}\hat{k}}_{tot}$. We will work under the assumption that the Killing vector is rotational, and the gravitational field is static. In this case, the orbital angular momentum integral yields
\beqa
L^{\hat{i}\hat{j}}&=& \frac{1}{k}\int_{S^2} \hat{r}^\alpha K^{\{\hat{i}\hat{j}\}\mu} \left( \utilde{\bep}_{0\mu,\alpha}-\underline{\bep}_{0\alpha,\mu}\right)\, r^2 d\Omega \\
  &=& -\frac{1}{k}\int_{S^2}\hat{r}^\alpha \lambda^{\{\hat{i}\hat{j}\}}_{\alpha\mu}  \left( 2\utilde{\bep}_0{}^\mu-\underline{\bep}_0 {}^\mu \right)\, r^2 d\Omega
\eeqa
and the spin contribution is 
\beq
S^{\hat{i}\hat{j}} =-\frac{1}{k}\int_{S^2}\hat{r}^\alpha \lambda^{\{\hat{i}\hat{j}\}}_{\alpha\mu}  \left( \utilde{\bep}_0{}^\mu+\underline{\bep}_0 {}^\mu \right)\, r^2 d\Omega 
\eeq
Adding these together we obtain
\beqa
J^{\hat{i}}_{tot}&=& \frac{1}{2}\epsilon^{\hat{i}}{}_{\hat{j}\hat{k}}(L^{\hat{i}\hat{j}}+S^{\hat{i}\hat{j}})=-\frac{3}{2k}\int_{S^2}\hat{r}^\alpha \,{\epsilon^{\hat{i}}}_{\hat{j}\hat{k}}\,\lambda^{\{\hat{j}\hat{k}\}}_{\alpha\mu} \,\utilde{\bep}_0{}^\mu \, r^2 d\Omega \\
&=& \frac{3}{8\pi G} \int_{S^2} \left(\bm{\hat{r}\times\mathcal{A}}\right)^{\hat{i}} \\
&=& \frac{3}{8\pi G} \int_{\Sigma} (\bm{\mathcal{B}})^{\hat{i}}\,.
\eeqa
From the form of the gravimagnetic field in the nonrelativistic limit (\ref{DipoleGField}),
\beq
J^{\hat{i}}_{tot}=-\alpha (\bm{L}+\bm{S})^{\hat{i}}\,.
\eeq
Using the standard convention $\alpha=-1$, this gives precisely the total (spin+orbital) angular momentum of the static current source.

\section{The Spin-Spin dipole interaction}
We can now partially answer one of the paradoxes discussed in the introduction. The well-known spin-orbit interaction as lauded by Mashhoon, Hehl and Ni, and others, gives a dipole-dipole interaction between the orbital angular momentum of a rotating matter distribution and the intrinsic spin of an elementary particle such as an electron or a neutrino. Opposite to the electromagnetic interaction, like poles attract. The force that the rotating mass distribution exerts on the intrinsic spin of the spinor is mediated by the well-understood frame-dragging effect of orbital angular momentum. On the other hand, if Newton's third law is to hold in the static Newtonian limit, the spinor must also exert an equal and opposite force on the rotating mass distribution. But what is the field that mediates this force? From the above discussion, the answer is now clear -- the intrinsic spin itself gives rise to a frame-dragging effect, which gives rise to the force that the spinor exerts on the rotating matter distribution. On the other hand, this opens up the possibility of an additional spin-spin interaction in addition to the well-known spin-orbit interaction. Furthermore, since the spin to mass ratio of fundamental particles is abnormally large, we expect the spin-spin interaction to be the dominant gravitational interaction at short range. We will now derive the weak field Hamiltonian and show that there is a characteristic spin-spin interaction term. 

\subsection{The weak field Hamiltonian}
We wish to derive the interaction Hamiltonian in a gauge invariant manner. Generically, due to diffeomorphism invariance, the total Hamiltonian in a spatially closed universe is identically zero. However, in the asymptotically flat context the Hamiltonian is not zero, and it is the additional boundary terms which give rise to the nonzero Hamiltonian. Under the assumption of a timelike Killing vector, appropriate in the static case, the boundary integral can be related to a bulk integral of the matter content, which in our case will give rise to the interaction Hamiltonian. In previous sections we have made the magnetostatic assumption which at the linearized level implies the existence of a timelike Killing vector. We will work with the covariant phase space consisting of solutions to the full set of equations of motion. As shown in \cite{RandonoSloan} the relation between the Hamiltonian as a boundary integral, and the bulk integral under the assumption of a timelike Killing vector is given by
\beqa
H_{tot}&=& \frac{1}{4k}\int_{S^2}\epsilon_{IJKL}\,\iota_{\bar{t}}(e^I\w e^J)\w \omega^{KL}\\
&=& \frac{\alpha}{6}\int_{\Sigma}\epsilon_{IJKL}e^I\w e^J\w e^K\,\frac{1}{2}\left(\bar{\psi}\gamma^L \Lie{t} \psi -\Lie{t}\bar{\psi} \,\gamma^L \psi \right)\\
&=& -\frac{\alpha}{2}\int_\Sigma \widetilde{\sigma}^{(3)}\,t_\mu \left(\bar{\psi}\gamma^\mu \Lie{t} \psi -\Lie{t}\bar{\psi} \,\gamma^\mu \psi \right)
\eeqa
where $ \widetilde{\sigma}^{(3)}$ is the induced metric volume on the three-surface given by $ \widetilde{\sigma}^{(3)}=\iota_{\bar{t}}(\frac{1}{4!}\epsilon_{IJKL}e^I\w e^J \w e^K \w e^L)$. Employing the on-shell Dirac equation this becomes
\beq
H_{tot}=-\frac{\alpha}{2}\int_{\Sigma}\widetilde{\sigma}^{(3)}\, {e^{a}}_I \left(\bar{\psi}\gamma^I\,D_a \psi -D_a\bar{\psi}\gamma^I \psi \right)-t_\nu {e^\nu}_I \bar{\psi} \{\gamma^I, \Gamma_0 \} \psi \,.
\eeq
In the linearized limit, this reduces to 
\beq
H_{tot}=H'_0+H^{(1)}+ H^{(2)}+H^{(3)}+H^{(4)}
\eeq
and we describe each term separately:
\begin{itemize}
\item{\underline{$H'_0$}:}    This is the free field Hamiltonian density with an overall modulating factor to account for the gravitational redshift giving rise to the Newtonian potential:
\beq
H'_0=\alpha \int_{\Sigma}d^3 x\ (1+\vep)\left(\frac{1}{2}(\bar{\psi}\gamma^a \p_a \psi -\p_a \bar{\psi} \gamma^a \psi) -m\bar{\psi}\psi \right)
\eeq
\item{\underline{$H^{(1)}$}:}    This term gives rise (in our choice of gauge) to the coupling of the orbital angular momentum of the particle to the external gravimagnetic field. In other words, it gives rise to the $\bm{\mathcal{B}}_{ext}\cdot \bm{L}$ coupling, and it is given explicitly by
\beq
H^{(1)}=-\alpha \int_\Sigma d^3 x\ \left(-\frac{1}{2} {\vep^a}{}_{\hat{0}}\left(\bar{\psi}\gamma^{\hat{0}} \p_a \psi -\p_a \bar{\psi} \gamma^{\hat{0}} \psi \right)\right)
\eeq
\item{\underline{$H^{(2)}$}:}    This term couples the external gravimagnetic field to the internal spin of the particle and gives rise (in our gauge) to the $\bm{\mathcal{B}}_{ext}\cdot \bm{S}$ coupling. It is
\beq
H^{(2)}=-\alpha \int_{\Sigma} d^3x\ \frac{1}{2}\bar{\psi} \{\gamma^0\,,\, \Gamma_0 \}\psi\,.
\eeq
\item{\underline{$H^{(3)}+H^{(4)}$}:}    These terms depend explicitly (in our gauge) only on ${\vep^a}_{\hat{i}}$ and therefore couple the pressure and shear of an external source to the spinor. In the gravi-magnetic approximation it is assumed that the energy-momentum and angular momentum will dominate over the pressure and shear terms. Thus we will neglect these terms. For completeness, they are given by
\beq
H^{(3)}+H^{(4)}=-\alpha \int_{\Sigma}d^3 x\ \left(-\frac{1}{2}\vep^a{}_i \left(\bar{\psi}\gamma^{\hat{i}} \p_a \psi -\p_a \bar{\psi} \gamma^{\hat{i}} \psi\right) +\frac{1}{2} \bar{\psi}\{\gamma^a\,,\,\Gamma_a\}\psi \right)\,.
\eeq  
\end{itemize}

To determine the precise form of the spin-orbit and spin-spin coupling, we make the following assumptions. Suppose we have two spinor sources $\psi_{1}$ and $\psi_{2}$ highly localized in regions $\mathcal{U}_1$ and $\mathcal{U}_2$, respectively, that are widely separated by an average distance $r$. The interaction Hamiltonian we are interested in comes from the terms $H^{(1)}+H^{(2)}$. We will assume that the linearized gravitational field coupled to $\psi_{1}$ is sourced entirely by $\psi_{2}$ and vice versa. Thus, we will ignore the self-interaction of the field. The symmetric components $\utilde{\bep}^I{}_\mu$ are determined in the linearized limit (in the Lorenz gauge) by the fermionic sources as shown above. The antisymmetric components are fixed by a choice for the internal $Spin(3,1)$ gauge. The calculations simplify if we choose a gauge such that $\bep_{\hat{0}a}=0$. With this gauge choice the tetrad perturbations takes the form
\beq
\bep_{\mu\nu}=
\left[\begin{array}{cc} 
\bep_{00} & \bep_{0b} \\
\bep_{a0} & \bep_{ab}
\end{array}\right]=
2G\left[\begin{array}{cc}
\left(\frac{Q^{\hat{0}}}{r}+\frac{\hat{r}^a p^{\hat{0}}_a}{r^2}\right) & 0 \\
-\frac{2 \hat{r}^c {M^{\hat{0}}}_{ca}}{r^2} & \frac{\hat{r}^c M_{acb}}{r^2}
\end{array}\right] \label{GaugedTetrad}
\eeq
where it is understood in this expression that $\hat{r}$ is the unit vector pointing from the source to the target and $r$ is the distance between the two as measured using the fiducial flat metric. 

Focus now on the first term in the Hamiltonian: 
\beq
H^{(1)}=-4G\alpha \int_{\mathcal{U}_1}d^3 x_1 \ \bm{m}_2 \cdot \left(\frac{\bm{\hat{r}_{21}}}{|\bm{r_{21}}|^3} \times \frac{1}{2}\left(\bar{\psi} \gamma^0\bm{\nabla} \psi -\bm{\nabla}\bar{\psi} \gamma^0 \psi \right)\right) \ +\ (1\leftrightarrow 2)
\eeq
where $\bm{r_{21}}=\bm{r_1}-\bm{r_2}$. Under the assumption that the regions $\mathcal{U}_1$ and $\mathcal{U}_2$ are small compared to the distance separating them, we can make the expansion in the region $\mathcal{U}_1$:
\beq
\frac{\bm{\hat{r}_{21}}}{|\bm{r_{21}}|^3} \approx \frac{\bm{r_1}-\bm{r_2}}{|\bm{r_2}|^3}-3\frac{\bm{\hat{r}_2}\,(\bm{\hat{r}_2}\cdot \bm{r_1})}{|\bm{r_2}|^3}\,.
\eeq
Using this approximation, and assuming the expectation value of the linear momentum of both particles is zero (which is necessary in the static case), after some manipulation the expression for $H^{(1)}$ reduces to
\beqa
H^{(1)}\approx-2G\alpha\ \frac{\left(3 (\bm{m_2 \cdot \hat{r}_2})(\bm{L_1 \cdot \hat{r}_2}) -\bm{m_2 \cdot L_1} \right)}{r^3} \ +\ (1\leftrightarrow 2)
\eeqa

We now turn to the second term in the Hamiltonian, $H^{(2)}$. From the identity $\{\gamma^I , \gamma^{[J} \gamma^{K]} \}=2i \epsilon^{IJKL}\gamma_5 \gamma_L$, and the linearized form of the spin connection, $\Gamma_{IJ\sigma}=\delta^{\mu\nu}_{IJ} \utilde{\vep}_{\sigma \mu,\nu}-\underline{\vep}_{IJ,\sigma}$, the expression reduces to
\beqa
H^{(2)}&=& -\alpha \int_{\mathcal{U}_1} d^3 x_1\  \frac{i}{4} \epsilon^{0ijk} \bar{\psi}_1 \gamma_5 \gamma_k \psi_1 \,\Gamma_{ij0} \  + \ (1\leftrightarrow 2)\\
&\approx & -\alpha \int_{\mathcal{U}_1} d^3 x_1\ \epsilon^{ijk} \partial_j \utilde{\vep}_{0i}  \left(\frac{-i}{2} \bar{\psi}_1 \gamma_5 \gamma_k \psi_1 \right)\  +\ (1\leftrightarrow 2) \\
&\approx & -\alpha \,\bm{\mathcal{B}}_2 \cdot \bm{S}_1\  +\ (1\leftrightarrow 2)
\eeqa
where $\bm{\mathcal{B}}_2$ is the gravimagnetic field produced by $\psi_2$, given by (\ref{DipoleGField}). Thus, in total we have
\beq
H^{(2)}\approx-2G\alpha\ \frac{\left(3 (\bm{m_2 \cdot \hat{r}_2})(\bm{S_1 \cdot \hat{r}_2}) -\bm{m_2 \cdot S_1} \right)}{r^3} +(1\leftrightarrow 2)\,.
\eeq

Putting both these terms together and inserting the form of the gravimagnetic moment $\bm{m}=-\frac{\alpha}{2}(\bm{L}+\bm{S})$, we have the total gravimagnetic interaction
\beqa
H^{(1)}+H^{(2)}&=&
2G\alpha^2 \  \frac{3 \left((\bm{L_1+S_1}) \cdot \bm{\hat{r}}\right) \left((\bm{L_2+S_2}) \cdot \bm{\hat{r}}\right) -(\bm{L_1+S_1}) \cdot (\bm{L_2+S_2})}{r^3} \nn\\
&=& 2G\alpha^2 \  \frac{3 \left(\bm{J_1}^{tot} \cdot \bm{\hat{r}}\right) \left(\bm{J_2}^{tot} \cdot \bm{\hat{r}}\right)-\bm{J_1}^{tot} \cdot \bm{J_2}^{tot}}{r^3}
\eeqa

In addition to the orbit-orbit and the spin-orbit terms, we have a new spin-spin interaction given by
\beq
H_{spin-spin}=2G\alpha^2 \  \frac{3 \left(\bm{S_1} \cdot \bm{\hat{r}}\right) \left(\bm{S_2} \cdot \bm{\hat{r}}\right) -\bm{S_1} \cdot \bm{S_2}}{r^3}\,.
\eeq
As expected, this interaction has the usual form of a dipole-dipole interaction, but in contrast to the electromagnetic interaction, north-pole attracts north-pole, and south-pole attracts south-pole. A similar interaction appears in the context of propagating torsion theories \cite{SabbataSivaram, Hammond--Torsion}, where the torsion mediates the interaction, as well as in the context of Maxwellian gravity \cite{BeheraNaik}. Perhaps because of the former result, the prevailing belief in the community appears to be that such a long-range interaction could not occur without propagating torsion, or could only be manifest as a contact interaction with nonpropagating torsion. In our case, the interaction comes from an ordinary gravitomagnetic effect derived from the coupling of tetrad gravity to a spinor field, and it is not mediated by torsion.

\section{When is the spin-spin interaction non-negligible?}
The Newtonian gravitational interaction between two fundamental particles is exceedingly small at large distances, and since the spin-spin force falls off like $\frac{1}{r^4}$, this force will be even smaller. However, as the particles come closer together eventually the magnitude of the spin-spin interaction will dominate over the ordinary attractive $\frac{1}{r^2}$ interaction. Fundamental particles have the peculiar property that their intrinsic spin to mass ratio is rather large, as compared to, for example, the extremal limit on the angular momentum of a Kerr black hole. Thus, the spin-spin interaction may dominate at distance scales much larger than one might ordinarily expect. We can make an heuristic argument to determine the distance scale at which the force becomes non-negligible. This will be true when the Hamiltonian for the ordinary Newtonian attractive force is of the same order of magnitude or less than the spin-spin interaction Hamiltonian. Using the minimum values of the interactions\footnote{Of course, in contrast to the $r^{-2}$ Newtonian force law, the $r^{-4}$ spin-spin force can be both positive and negative, and therefore has the potential to cancel itself out at the microscopic level. By ``dominate" interaction here, we simply mean the interaction which we can never justifiably neglect without additional assumptions.}, the spin-spin interaction will dominate over the Newtonian interaction within the range given by 
\beq
\Big|G\frac{m_1 m_2}{r} \Big| \lesssim \Big| 4 G\frac{s_1 s_2}{r^3} \Big|
\eeq 
or, including the factor of c,
\beq
r^2 \lesssim  \Big| 4 \frac{s_1}{m_1 c} \frac{s_2}{m_2 c} \Big|\,.
\eeq
For spinors the maximum magnitude of the spin is $s=\frac{\hbar}{2}$, so we obtain
\beq
r \lesssim \sqrt{\lambda^{(1)}_c \lambda^{(2)}_c}
\eeq
where $\lambda^{(1)}_c=\frac{\hbar}{m_1 c}$ and $\lambda^{(2)}_c=\frac{\hbar}{m_2 c}$ are the (reduced) Compton wavelengths of the interacting particles. Using the explicit form for the tetrad perturbations, $\ref{GaugedTetrad}$, it is clear that even at these length scales we have $|\bep_{\mu\nu}|\ll 1$, so the linearized approximation is still valid near the Compton scale. Thus the spin-spin interaction is expected to be the dominant interaction between fundamental particles at length scales smaller than the geometric mean of the Compton wavelengths of the interacting particles. 

The spin-spin interaction involves the internal angular momentum, which also couples to the electromagnetic field to yield a dipole-dipole coupling. Since the electromagnetic interaction is much stronger than the gravitational interaction, the effect will be most prominent in the dynamics of weakly interacting neutral particles\footnote{Strongly interacting particles are likely to be poor candidates for phenomenology of this interaction because they are typically composite particles made of electrically charged quarks. For example, despite being electrically neutral, the neutron has a small ordinary magnetic moment coming from its composite nature. Despite being small, it is likely that the electromagnetic spin-spin interaction between neutrons would dominate over the gravitational spin-spin interaction at all length scales.}. Taking the upper limit on the lightest neutrino to be around $1\, eV$, the scale at which the spin-spin interaction dominates is around $10^{-5} \ cm$. At these length scales, we are well within the bounds of validity of the linearized theory. On the other hand, the Compton wavelength effectively sets the scale at which quantum field theory effects become non-negligible since the Compton wavelength is the wavelength a photon would need to create a particle of mass $m$ out of the vacuum. Evidently, the regime where the spin-spin interaction becomes significant coincides with the regime where quantum field theoretical effects become appreciable. Thus, we propose the following ``rule of thumb":
\begin{quotation}
{\noindent \it In any process of sufficiently high energy for quantum field theoretical effects to be appreciable in the interaction of two spin-$\frac{1}{2}$ particles, the gravitationally modulated spin-spin interaction is expected to be of equal or greater significance than the ordinary attractive Newtonian gravitational interaction.}
\end{quotation}
Alternatively, one could say that quantum field theoretical effects are likely to be the most prominent in the dominant interaction in the regime where quantum processes are non-negligible. Following this line of reasoning one may conclude that quantum effects should be the most prominent in the spin-spin interaction. To support this argument we can compare the situation with that of quantum electrodynamics: In classical electrodynamics, the scale at which the spin- spin interaction potential $\mu_0(\bm{\mu_1}\cdot \bm{\mu_2}-3\bm{\mu_1}\cdot \bm{r}\,\bm{\mu_2}\cdot \bm{r})/ 4\pi r^3$ dominates over the charge potential $q_1 q_2/{{4\pi \epsilon_0}r}$ is also classically given by the geometric mean $r\lesssim\sqrt{\lambda^{(1)}_{c} \lambda^{(2)}_{c}}$. In light of this, it is not surprising that the most prominent quantum effects of QED are manifest in the spin-spin interaction, namely in the correction of the gyromagnetic ratio. Although quantum gravity effects are expected to be dramatically smaller, following this same line of reasoning one might expect that some of the most prominent effects of perturbative quantum gravity or semiclassical gravity with backreaction considerations will involve the spin-spin interaction.

As a final comment, we note that the primary purpose of this paper is to demonstrate the existence of a generic spin-spin interaction. It remains to be seen if this interaction has any application to the standard model of particle theory, cosmology, or astrophysics. To answer these questions, a more realistic model is necessary. Such a model would necessarily take into account the effects of the weak and strong interaction, and since these interactions are generally much stronger than the gravitational interaction, and sine each has its own spin-spin interactions associated with it, it may seem that weak and strong effects will override the spin-spin gravitational interaction in the regime where it begins to dominate over the Newtonian potential. However, a simple heuristic argument shows that there may be a window where the gravitational spin-spin interaction may be the dominant, albeit small, interaction. As a simple model consider the wave equation for a massive vector boson (which can be used to model the gauge bosons of the weak force in a linearized approximation):
\beq
(\Box-(\mu c /\hbar)^2) A^\nu=-\kappa J^\nu
\eeq
where $\mu$ is the mass of the gauge field. As shown previously, using a retarded Green's function this relation can be inverted to give the Yukawa vector potential
\beq
A^\nu(x)=\frac{\kappa}{4\pi}\int_{\Lambda_{-}}d^3x'\ \frac{e^{-\mid\bm{r}-\bm{r'} \mid/\lambda_\mu} \,J^\nu(x')}{\mid\bm{r}-\bm{r'} \mid}
\eeq
where $\lambda_{\mu}=\frac{\hbar}{\mu c}$ is the Compton wavelength of the massive boson. From this it is clear that the weak effects of a massive gauge mediated force fall off rapidly outside of the scale set by the Compton wavelength, $r\sim \lambda_\mu$. For weakly interacting particles, such as neutrinos, the gauge bosons have a mass of around $100 \,GeV$ corresponding to a length scale around $10^{-18}\,m$. As we have seen, the spin-spin interaction for particles with masses around $1\,eV$ dominates at around $10^{-7}\,m$. So there is a large window where the spin-spin interaction could potentially dominate over the Newtonian interaction and not be washed out by electroweak phenomena. It should be stressed that this is a heuristic model only, and more accurate predictions would require a neutrino mass extension of the standard model together with semiclassical gravitational couplings. More thorough investigations of this sort are currently in progress. 

\section{Concluding Remarks}
In this paper we have investigated the gravitational effects of a Dirac field as a source of matter, focusing on the effect of the internal spin-angular momentum on the spacetime geometry. We have shown that in the context of the linearized theory, despite the absence of a physical rotating mass distribution, the intrinsic spin of an Dirac field gives rise to a frame-dragging effect precisely analogous to that of ordinary orbital angular momentum. This resolves some of the paradoxes mentioned in the introduction. In particular, in the Newtonian limit, Newton's third law is regained for the spin-orbit interaction because the orbiting mass experiences a force mediated by the gravimagnetic field produced by the spinor itself. This yields the surprising consequence that with two spinor sources, the gravimagnetic field mediates a long-range spin-spin interaction analogous to the dipole-dipole interaction of ordinary electromagnetism. It remains to be seen if such an interaction has any astrophysical or cosmological consequences. The effect will be most pronounced in the interaction of a neutral particle with another spin source since otherwise the corresponding electromagnetic interaction will dominate over the gravitational interaction. Thus, we expect that such an interaction may be significant in the dynamics of neutrinos as they interact with other spinor fields. The consequences of such interactions are currently under investigation.

\section*{Acknowledgments} I would like to thank Bahram Mashhoon, Friedrich Hehl, Yuri Obukhov, Chris Beetle, and Tom\'{a}\v{s} Liko for comments and discussions as well as for directing me toward many references. This research was supported in part by NSF Grant No. OISE0853116, NSF Grant No. PHY0854743, The George A. and Margaret M. Downsbrough Endowment, and the Eberly Research 
Funds of Penn State.

\bibliography{MasterBibDesk}

\end{document}